\magnification=\magstep1
\input aua.mac
\def\artanh{\mathrel{\rm arctanh}}
\MAINTITLE{Arcs from a Universal Dark-Matter Halo Profile}
\AUTHOR{Matthias Bartelmann}
\INSTITUTE{Max-Planck-Institut f\"ur Astrophysik, P.O. Box 1523,
D--85740 Garching, Germany}
\ABSTRACT{Navarro, Frenk, \& White have recently found numerically
that the density profile of dark-matter halos can be described by a
universal two-parameter function over a broad range of halo
masses. The profile is singular, approaching the halo center with
$\rho\propto r^{-1}$. It had been argued previously that radially
distorted, gravitationally lensed images of background sources in
galaxy clusters, so-called radial arcs, required a flat core in the
cluster density profile. Such radial arcs have so far been detected in
two galaxy clusters, in apparent contradiction with a singular density
profile. I show here that the profile suggested by Navarro et al. can
produce radial arcs despite its central singularity, and describe how
the two parameters of the profile can be determined in clusters where
radial and tangential arcs are observed. I then apply this analysis to
the two clusters where radial arcs were detected. In both cases, the
redshifts of the radial arcs are yet unknown, hence definitive
conclusions on the profile parameters cannot yet be drawn. Numerically
determined values for the parameters of cluster-sized halos can,
however, be used to predict the range of the unknown arc redshifts,
thus providing a direct observational test for the proposed density
profile. A potential difficulty with the profile is that the radial
magnification of tangential arcs is large, hence tangential arcs
should be thick or their sources should be very thin in the radial
direction.}
%
\titlea{Introduction}
Navarro, Frenk, \& White (1995a,b) have recently found that the
density profile of dark-matter halos numerically simulated in the
framework of the standard CDM cosmogony can very well be described by
the radial function
$$
  \rho(x) = {\rho_{\rm s}\over x(1+x)^2}\;,
\eqno(1.1)$$
within the broad halo mass range $3\times10^{11}\la
M_{200}/M_\odot\la3\times10^{15}$. Particular care was taken to
investigate and rule out numerical artifacts of the simulations. The
radial coordinate $x$ is the radius in units of a scale radius $r_{\rm
s}$, $x\equiv r/r_{\rm s}$. The profile flattens towards the halo
center but does not have a flat core. Although Navarro et
al. restricted their original study to the standard CDM cosmogony, it
has turned out that (1.1) provides excellent fits also to halos formed
in other cosmogonies or from density fluctuations described by a
variety of power-law perturbation spectra (Cole \& Lacey 1995). It
therefore appears that the profile (1.1) can be considered as a
universal two-parameter function describing the structure of
dark-matter halos, independent of the size of the halos and the
environment in which they form.

On the other hand, it has frequently been stated before that the
occurrence of so-called radial arcs in galaxy clusters required the
density to be constant approximately out to the distance of the radial
arc from the center of the cluster, and core radii have been derived
from the positions of radial arcs (e.g. Mellier, Fort, \& Kneib 1993;
Smail et al.~1995). This statement was mainly supported by one
particular choice for the density profile of clusters, the
non-singular isothermal profile, for which the location of radial arcs
indeed provides an estimate for the core radius. Noting that profiles
with a flat core may be poor choices for real clusters,
Miralda-Escud\'e (1995) has recently studied a variety of singular and
non-singular density profiles with regard to the location of radial
arcs and their counter-images. Radial arcs are gravitationally
distorted images of galaxies in the background of clusters which are
radially elongated rather than tangentially as most giant luminous
arcs are. They are much less numerous than (tangential) giant luminous
arcs. So far, only in the clusters MS~2137 (Fort et al.~1992) and
Abell~370 (Smail et al.~1995) have radial arcs been reported, but
several more have been observed (J.-P. Kneib 1996, private
communication). In both cases, the separation of the radial arcs from
the cluster centers are such that they were interpreted in terms of
cluster core sizes on the order of $20\,h^{-1}$kpc, where the Hubble
constant is $H_0=100\,h$ km s$^{-1}$ Mpc$^{-1}$. The implications of
the radial arc in MS~2137 for a variety of mass profiles have also
been discussed by Miralda-Escud\'e (1995).

I show in Sect.~2 of this paper that the singular profile (1.1)
necessarily produces a radial critical curve. I also show that the
existence of the radial critical curve is stable against perturbing
the lens with an external shear field as long as the shear is small,
$|\gamma|<1$. I describe how the two parameters of the profile (1.1)
can be measured in clusters where radial and tangential arcs are
observed. In Sect.~3, I analyze the two cases of radial arcs so far
detected, and in Sect.~4 I summarize and discuss the results.
\titlea{Occurrence of Radial Arcs}
\titleb{Existence of Radial Critical Curves}
Axially symmetric lenses are completely described by their
surface-mass density $\Sigma(x)$ in units of a critical surface mass
density $\Sigma_{\rm cr}$,
$$
  \kappa(x) \equiv {\Sigma(x)\over\Sigma_{\rm cr}}\;,
\eqno(2.1)$$
where $\Sigma_{\rm cr}$ is determined by the angular-diameter
distances between the observer and the lens $D_{\rm d}$, the lens and
the source $D_{\rm ds}$, and the observer and the source $D_{\rm s}$,
$$
  \Sigma_{\rm cr} = {c^2\over4\pi G}\,
  {D_{\rm s}\over D_{\rm d}D_{\rm ds}}\;.
\eqno(2.2)$$
For general references on gravitational lensing, see Schneider,
Ehlers, \& Falco (1992) or Blandford \& Narayan (1992).

The mass inside radius $x$ is conveniently described by the
dimensionless function
$$
  m(x) \equiv 2\int_0^x\,dy\,y\kappa(y)\;.
\eqno(2.3)$$
Locally, the lens mapping is described by its Jacobian matrix ${\cal
A}$, which is symmetric and thus can be diagonalized. Its two
eigenvalues are
$$
  \lambda_{\rm r} = 1 - {d\over dx}\,{m\over x}\;,\quad
  \lambda_{\rm t} = 1 - {m\over x^2}\;.
\eqno(2.4)$$
Radial and tangential critical curves arise if and when the conditions
$$
  \lambda_{\rm r} = 0\,,\quad\lambda_{\rm t} = 0
\eqno(2.5)$$
are satisfied, respectively.

The density profile (1.1) implies the surface mass density
$$
  \Sigma(x) = {2\rho_{\rm s}r_{\rm s}\over x^2 - 1}\,f(x)\;,
\eqno(2.6)$$
with
$$
  f(x) = \cases{
  1 - {2\over\sqrt{x^2 - 1}}\arctan\sqrt{x-1\over x+1} & $(x>1)$ \cr
  1 - {2\over\sqrt{1 - x^2}}\artanh\sqrt{1-x\over 1+x} & $(x<1)$ \cr
  0 & $(x=1)$ \cr
  }\;.
\eqno(2.7)$$
If we define $\kappa_{\rm s}\equiv\rho_{\rm s}r_{\rm s}\Sigma_{\rm
cr}^{-1}$,
$$
  \kappa(x) = 2\kappa_{\rm s}\,{f(x)\over x^2-1}\;,
\eqno(2.8)$$
and the dimensionless mass $m(x)$ becomes
$$
  m(x) = 4\kappa_{\rm s}\,g(x)\;,
\eqno(2.9)$$
where
$$
  g(x) = \ln{x\over2} + \cases{
  {2\over\sqrt{x^2-1}}\arctan\sqrt{x-1\over x+1} & $(x>1)$ \cr
  {2\over\sqrt{1-x^2}}\artanh\sqrt{1-x\over 1+x} & $(x<1)$ \cr
  1 & $(x=1)$ \cr
  }\;.
\eqno(2.10)$$
It is now easy to see that
$$\eqalign{
  {d\over dx}{m\over x}&\to\infty \quad(x\to0)\cr
  {d\over dx}{m\over x}&\to0 \quad(x\to\infty)\cr}\;.
\eqno(2.11)$$
Since $(d/dx)(m/x)$ is continuous, it follows from eq. (2.11) that
there must be a radius $x_{\rm r}$ where $\lambda_{\rm r}=0$ is
satisfied, and hence the singular profile (1.1) must have a radial
critical curve for any value of $\rho_{\rm s}$ and $r_{\rm
s}$. Therefore, despite of its central singularity, the density
profile (1.1) can produce radial arcs. Needless to say, the profile
(1.1) also always has a tangential critical curve at radius $x_{\rm
t}$. Figure 1 displays the radii $x_{\rm r}$ and $x_{\rm t}$, the
radial eigenvalue at the position of the tangential critical curve
$\lambda_{\rm r}(x_{\rm t})$ and the tangential eigenvalue at the
position of the radial critical curve $\lambda_{\rm t}(x_{\rm r})$ as
a function of $\kappa_{\rm s}$.

\begfigside{fig1.tps}
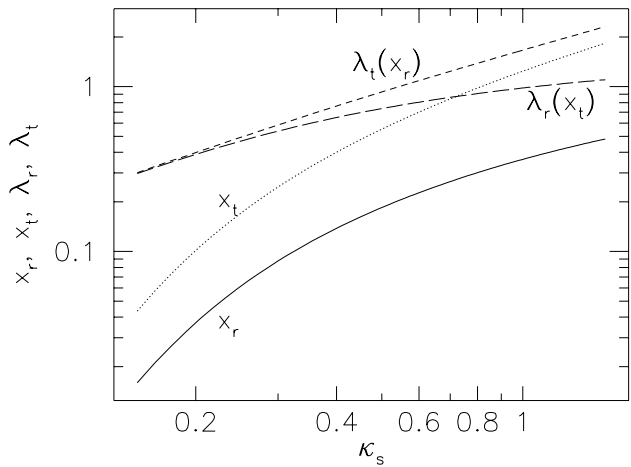
\figure{1}{Locations $x_{\rm r,t}$ of the radial and the tangential
critical curve (solid and dotted lines, respectively) of the density
profile (1.1), and radial and tangential eigenvalues $\lambda_{\rm
r}(x_{\rm t})$ and $\lambda_{\rm t}(x_{\rm r})$ at the positions of
the tangential and radial critical curves (long- and short-dashed
lines, respectively), all as functions of $\kappa_{\rm s}$.}
\endfig

\titleb{Influence of External Shear}
In the presence of external shear $\gamma>0$, the eigenvalues of the
Jacobian matrix of the lens mapping are changed to
$$
  \lambda_{\rm r}\to\lambda_{\rm r}+\gamma\cos2\theta\;,\quad
  \lambda_{\rm t}\to\lambda_{\rm t}-\gamma\cos2\theta\;,
\eqno(2.12)$$
where $\theta$ is the polar angle. These expressions are valid to
first order in $\gamma$. Coordinates have been chosen such that the
coordinate axes are the principal axes of the external shear
matrix. This can be done without loss of generality because the
unperturbed profile is axially symmetric.

We first note that the condition for the occurrence of a radial
critical curve remains unchanged by a spatially constant external
shear field for $\gamma<1$. From the condition $\lambda_{\rm r}=0$
and eq. (2.4), we have
$$
  {d\over dx}\;{m\over x} = 1+\gamma\cos2\theta\;.
\eqno(2.13)$$
We have seen in eq. (2.11) that the left-hand side takes any positive
value, so that the presence of an external shear field $\gamma<1$
changes the location of the radial critical curve, but cannot make it
disappear.

The amount by which the critical curves are moved by the external
shear field follows from
$$
  \Delta x = \left({d\det{\cal A}\over dx}\right)^{-1}\,
  \Delta\det{\cal A}\;,
\eqno(2.14)$$
where ${\cal A}$ is the Jacobian matrix of the lens mapping and
$\det{\cal A}=\lambda_{\rm r}\lambda_{\rm t}$. To first order in
$\gamma$, we find from eq. (2.12) that external shear changes
$\det{\cal A}$ by
$$
  \Delta\det{\cal A} = \gamma\cos2\theta\,
  (\lambda_{\rm t}-\lambda_{\rm r})\;,
\eqno(2.15)$$
and eq. (2.14) becomes
$$
  \Delta x = \gamma\cos2\theta\,{(\lambda_{\rm t}-\lambda_{\rm r})
  \over
  \lambda_{\rm t}'\lambda_{\rm r}+\lambda_{\rm t}\lambda_{\rm r}'}\;,
\eqno(2.16)$$
where primes denote the derivative with respect to $x$. Hence, the
radial and the tangential critical curves (where $\lambda_{\rm r}=0$
and $\lambda_{\rm t}=0$, respectively) are shifted by
$$
  \Delta x_{\rm r} =
  {\gamma\cos2\theta\over\lambda_{\rm r}'(x_{\rm r})}\;,\quad
  \Delta x_{\rm t} =
  -{\gamma\cos2\theta\over\lambda_{\rm t}'(x_{\rm t})}\;.
\eqno(2.17)$$
Since the derivatives of $\lambda_{\rm r}$ and $\lambda_{\rm t}$ at
the locations of the respective critical curves are of order unity
(cf. Fig.~2), the shifts of the critical curves are of order of the
external shear $\gamma$. We shall see later that the conclusions
obtained from the radial and tangential arcs in MS~2137 and A~370 do
not depend sensitively on the precise locations of the arcs. Therefore
a possible shift of the critical curves by an external shear field is
insignificant in the cases of these clusters. Also, the mass
contributed by a central cluster galaxy can shift in particular the
radial critical curve (Miralda-Escud\'e 1995), but it would not alter
the conclusions below for the same reason as for an external shear
field. I therefore neglect a possible contribution from a central
galaxy altogether.

\begfigside{fig2.tps}
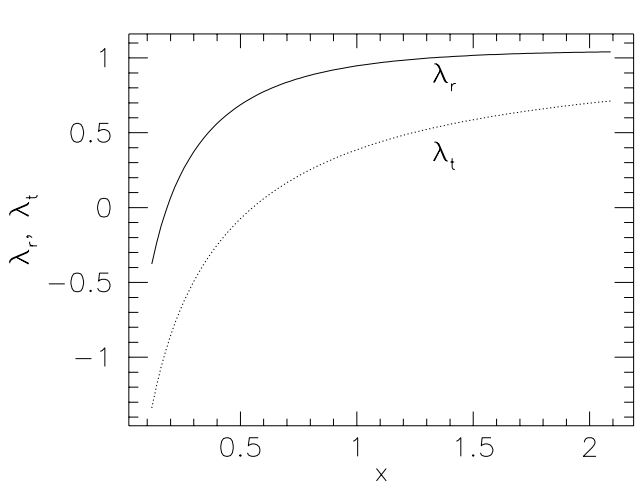
\figure{2}{Radial and tangential eigenvalues $\lambda_{\rm r}$ and
$\lambda_{\rm t}$ (solid and dotted lines, respectively) as a function
of radial distance $x$, for $\kappa_{\rm s}=0.5$.}
\endfig

\titleb{Profile Parameters from Tangential and Radial Arcs}
If radial and tangential arcs appear in one cluster, we can infer the
parameters $r_{\rm s}$ and $\rho_{\rm s}$ from the location of the two
arcs alone.

Radial and tangential arcs appear close to the respective critical
curves. Therefore their positions $x_{\rm r}$ and $x_{\rm t}$ must
satisfy eq. (2.5). From eq. (2.4), we obtain for the position of the
tangential arc
$$
  4\kappa_{\rm s,t}{g(x_{\rm t})\over x_{\rm t}^2} = 1\;,
\eqno(2.18)$$
and from the position of radial arc we can conclude
$$
  4\kappa_{\rm s,r}\left(
  {g'(x_{\rm r})\over x_{\rm r}}-{g(x_{\rm r})\over x_{\rm r}^2}
  \right) = 1\;.
\eqno(2.19)$$
Note that the factor $\kappa_{\rm s}$ is generally different for the
two arcs because their sources can be at different redshifts.

Combining eqs. (2.18) and (2.19), and taking eq. (2.2) into account,
we have
$$
  \sigma \equiv {\Sigma_{\rm cr,r}\over\Sigma_{\rm cr,t}} =
  \left({g'(\alpha x_{\rm t})\over\alpha x_{\rm t}} -
        {g(\alpha x_{\rm t})\over(\alpha x_{\rm t})^2}\right)
  {x_{\rm t}^2\over g(x_{\rm t})}\;,
\eqno(2.20)$$
where $\alpha$ is defined as
$$
  \alpha \equiv {x_{\rm r}\over x_{\rm t}}\;.
\eqno(2.21)$$
The ratio $\sigma$ between the critical surface mass densities for the
radial and the tangential arc depends on the redshifts of the arcs and
the cluster, and on the cosmological parameters. For definiteness, we
assume an Einstein-de Sitter model universe in the following where
necessary. The ratio $\alpha$ is observable, and so is the ratio
$\sigma$, at least in principle. Given $\alpha$ and $\sigma$, we can
solve eq. (2.20) for $x_{\rm t}$. Since we know the physical distance
of the tangential arc from the cluster center $r_{\rm t}$, we can then
infer the scale radius $r_{\rm s}$ of the cluster profile, and either
of eqs. (2.18) and (2.19) then yields $\rho_{\rm s}$. In practice, the
problem is of course that the redshifts of the arcs are difficult to
measure because of their generally low surface brightness. I shall now
discuss the two known cases where radial and tangential arcs have been
found in galaxy clusters.
\titlea{The Cases of MS~2137 and A~370}
\titleb{MS~2137}
The first radial arc was detected by Fort et al. (1992) in the cluster
of galaxies MS~2137, which is at redshift $z_{\rm c}=0.315$. In the
same cluster, there is an approximately tangentially oriented, giant
luminous arc. The radial arc is at $r_{\rm r}=5\farcs0$ from the
cluster center, while the tangential arc is at $r_{\rm
t}=15\farcs5$. Redshift information is not available for either arc,
but Mellier, Fort, \& Kneib (1993) tentatively conclude from the color
of the tangential arc that its redshift is $z_{\rm
t}=1.5\pm0.5$. Hence for MS~2137, $\alpha=0.32$, but $\sigma$ is
unknown. In Fig.~3, I plot $x_{\rm t}$ as a function of $\sigma$ for
$\alpha=0.32$ and for $0.75\alpha$ and $1.25\alpha$.

\begfigside{fig3.tps}
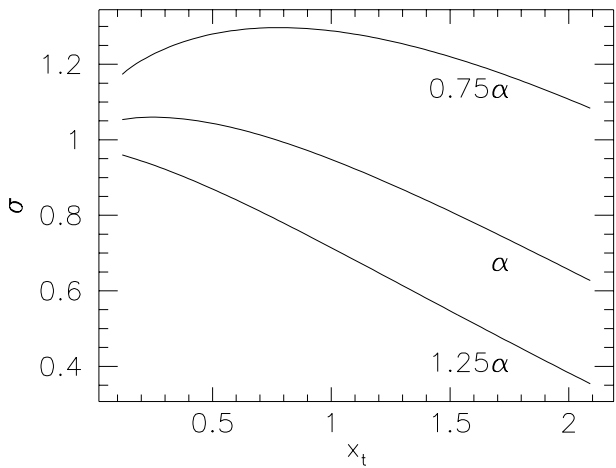
\figure{3}{$\sigma$ as function of $x_{\rm t}$ for $\alpha=0.32$
(central line) and for $\alpha$ lowered and increased by 25\%, as
indicated.}
\endfig

Since the redshifts of the arcs are unknown, $\sigma$ is so far
unconstrained, and we therefore cannot infer $r_{\rm s}$ and
$\rho_{\rm s}$ for the cluster MS~2137. Numerical simulations of dark
halos within the framework of the CDM cosmogony predict that $r_{\rm
s}$ is approximately $20\%$ of the tidal radius $r_{200}$, or on the
order of $\sim250\,h^{-1}$ kpc for a cluster-mass halo. The tangential
arc in MS~2137 is at $r_{\rm t}=15\farcs5$, or
$$
  r_{\rm t} = 45\,h^{-1}\;{\rm kpc}
\eqno(3.1)$$
at the cluster redshift. This implies $x_{\rm t}\sim0.2$. We can read
off from Fig.~3 that such low values of $x_{\rm t}$ indicate
$1\la\sigma\la1.2$. To see what this means for the arc redshifts, I
plot in Fig.~4 $\sigma$ as a function of the radial-arc redshift
$z_{\rm r}$ for five values of the tangential-arc redshift $z_{\rm
t}$.

\begfigside{fig4.tps}
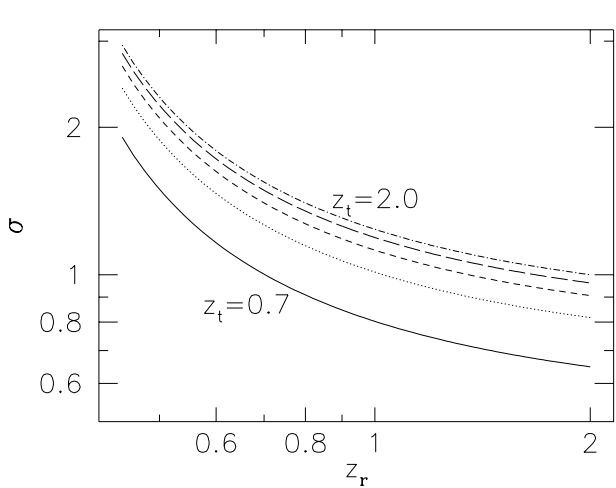
\figure{4}{$\sigma$ as a function of the radial-arc redshift $z_{\rm
r}$ for the five tangential-arc redshifts $z_{\rm
t}\in\{0.7,1.0,1.4,1.7,2.0\}$.}
\endfig

The figure shows that $\sigma$ is monotonically decreasing for
increasing $z_{\rm r}$. By definition, it is unity if both arcs are at
the same redshifts. Hence the requirement that $\sigma\simeq1$ from
above predicts that, if the cluster MS~2137 can be modeled with the
density profile (1.1), the arcs either have to be at very similar
redshifts, or that both have to be at much higher redshifts than the
cluster, $z_{\rm r,t}\ga1$. If spectroscopy should reveal that the
arcs are at discrepant redshifts, it would mean that the cluster MS
2137 cannot be modeled with the profile (1.1).

There is, however, another difficulty with $x_{\rm t}$ being so low as
$\sim0.2$, which can be read off from Fig.~1. Consider the radial
eigenvalue $\lambda_{\rm r}(x_{\rm t})$ at the position of the
tangential critical curve. The radial magnification of the tangential
arc is given by the inverse of $\lambda_{\rm r}(x_{\rm t})$, which is
plotted for $\kappa_{\rm s}=0.5$ in Fig.~2. Since for $x_{\rm
t}\sim0.2$ the radial eigenvalue is $\lambda_{\rm r}(x_{\rm
t})\sim0.5$, the tangential arc should be magnified in the radial
direction by a factor of about two. Mellier et al. (1993) report that
the arc seems embedded in a faint halo of width $2''$, and the main
blue bright structure is marginally resolved or unresolved at a seeing
of $0\farcs8$. The source should therefore have a radial width of
$\la0\farcs4$, or $\la1''$ for the more extended halo. Thin tangential
arcs located well within the scale radius $r_{\rm s}$ therefore
require sources which are very narrow in the radial direction.

Generally, the width magnification of a tangential arc is in good
approximation given by to $[2(1-\kappa(x_{\rm t}))]^{-1}$ (e.g. Hammer
1991), where $\kappa(x_{\rm t})$ is the scaled surface-mass density at
the location of the tangential arc. Statistically, cluster
substructure tends to increase the separation of tangential arcs from
the cluster center. Then, $\kappa(x_{\rm t})$ and therefore also the
radial magnification are decreased (Bartelmann, Steinmetz, \& Weiss
1995). The difficulty with the large width magnifications of
tangential arcs implied by the profile (1.1) is therefore alleviated
in the presence of cluster substructure.
\titleb{A~370}
The radial arc in Abell~370 was reported by Smail et al. (1995), while
the giant tangential arc in that cluster was the first arc to be
discovered (Soucail et al.~1987; Lynds \& Petrosian 1989). The
situation is more complicated in this cluster because it is evidently
bimodal. If we assume, however, that the cluster clump where both the
tangential and the radial arcs are found can be modeled with the
profile (1.1), we can apply a similar analysis to that cluster
also. The cluster is at redshift $z_{\rm c}=0.375$, and the ratio
between the distances of the radial and the tangential arc to the
center of the subcluster where they are found is approximately
$\alpha=0.7$ (read off from Fig.~2 of Smail et al.~1995). The angular
separation between the tangential arc and the subcluster center is
$r_{\rm t}\simeq10''$, or
$$
  r_{\rm t} \simeq 32\,h^{-1}\;{\rm kpc}
\eqno(3.2)$$
at the redshift of the cluster. As in Fig.~3, we plot in Fig.~5
$\sigma$ as a function of $x_{\rm t}$ for $\alpha$ and
$0.75\alpha,1.25\alpha$.

\begfigside{fig5.tps}
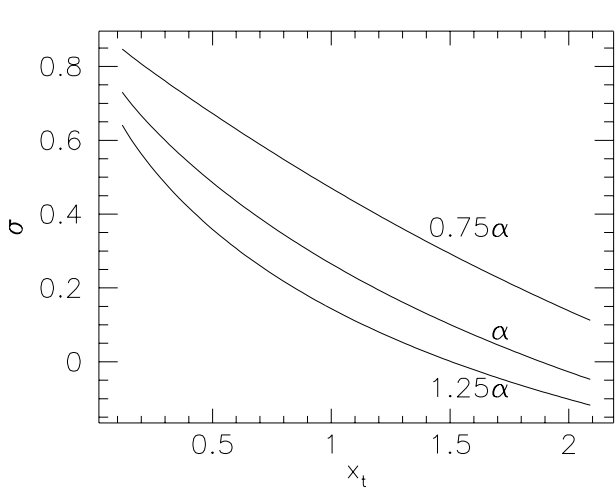
\figure{5}{$\sigma$ as function of $x_{\rm t}$ for $\alpha=0.7$
(central line) and for $\alpha$ lowered and increased by 25\%, as
indicated.}
\endfig

First, it is reassuring that the curves in Fig.~5 for the three values
of $\alpha$ are not much different for $x_{\rm t}\la0.5$. We have
discussed above that an external shear field $\gamma$ can shift the
critical curves and thus change $\alpha$ by an amount of order
$\gamma$. Therefore, a possible shift of the critical curves by the
shear field expected from the second subcluster in A~370 does not have
a pronounced influence. Again, the redshift of the radial arc is
unknown so that $\sigma$ remains unconstrained. Assuming again the
results from numerically simulated cluster-mass halos, we infer that
$x_{\rm t}\simeq0.13$, and then $0.6\la\sigma\la0.8$. From Fig.~4, we
can then predict that the arc redshifts cannot be the same. Since
$\sigma<1$, the radial arc must be at a higher redshift than the
tangential arc. The profile (1.1) requires the tangential arc to be at
a redshift $z_{\rm t}\la0.7$, which is in good qualitative agreement
with the redshift $z_{\rm t}=0.724$ spectroscopically determined by
Soucail et al. (1988). The radial arc must then be at a very high
redshift, $z_{\rm r}\sim1.5$. Again, if the radial arc redshift should
be lower than $z_{\rm r}\sim1$, the subcluster of A~370 where the arcs
are located cannot be described by the profile (1.1). From the
detailed and successful lens model for A~370 by Kneib et al. (1993),
Smail et al. (1995) conclude that the radial arc must be at a redshift
$z_{\rm r}=1.3\pm0.2$, again in qualitative agreement with the
prediction above. The same difficulty with the radial magnification
of the tangential arc as in MS~2137 applies here. If $x_{\rm
t}\sim0.13$ as the numerical simulations indicate, then the radial
magnification should be $\sim2.5$, and then the radial intrinsic
source size should be $\sim0\farcs6$. Again, cluster substructure can
reduce the width of tangential arcs as discussed before.

From Fig.~1, we can read off that $x_{\rm t}\sim0.13$ corresponds to
$\kappa_{\rm s}\sim0.2$. For the redshifts of A~370 and the tangential
arc in that cluster, eq. (2.2) implies $\Sigma_{\rm cr}=1.41\,h$ g
cm$^{-2}$, hence $\Sigma_{\rm s}\equiv\rho_{\rm s}r_{\rm
s}\sim0.28\,h$ g cm$^{-2}$. With $r_{\rm s}\sim250\,h^{-1}$ kpc, this
yields for the characteristic overdensity defined by Navarro et
al. (1995b)
$$
  \delta_{\rm c}\sim2\times10^{4}
\eqno(3.3)$$
for $\Omega_0=1$, corresponding to a concentration parameter (Navarro
et al.~1995b) of
$$
  c \equiv {r_{200}\over r_{\rm s}}\sim7\,.
\eqno(3.4)$$
This is again in good qualitative agreement with the results from
numerical simulations.
\titlea{Summary and Conclusions}
Navarro et al. (1995a,b) found that the dark-matter density profiles
of halos formed in the CDM cosmogony can very well be fit with the
two-parameter function (1.1) over a very wide range of halo
masses. Simulations by Cole \& Lacey (1995) confirm this result for a
variety of power-law density perturbation spectra. Here, I have
investigated whether this particular profile, which seems to be a
universal function describing the structure of dark-matter halos
independent of their size and environment, can account for the
radial arcs observed in two clusters of galaxies despite the claim
that radial arcs require non-singular, flat density profiles. I showed
that the profile necessarily produces a radial critical curve for any
combination of the two parameters $\rho_{\rm s}$ and $r_{\rm s}$, and
that the result is robust against perturbing the potential with an
external shear field $\gamma<1$.

For clusters which produce radial and tangential arcs simultaneously,
both profile parameters can be inferred if the distances of the arcs
from the cluster center and their redshifts can be measured. Radial
arcs have so far been reported in two clusters only, viz. MS~2137 and
A~370, but some more have recently been observed. It is of potential
importance for cosmological models to measure the parameters of the
potential (1.1) because numerical simulations indicate that they
depend on cosmological parameters.

In MS~2137, the redshifts of both arcs are unknown. Assuming scale
radii on the order of $250\,h^{-1}$ kpc which are found in numerical
simulations, I have shown that the ratio between the angular
separations of the radial and the tangential arc from the cluster
center in MS~2137 requires them to be either at very similar
redshifts, or at redshifts $\ga1$. In A~370, the same reasoning leads
to the requirement that the tangential arc must be at a redshift
$\la0.7$ and the radial arc at a redshift $\sim1.5$ in order to
reconcile their relative positions with the profile (1.1) and the
numerically found scale radius. Spectroscopy (Soucail et al.~1988)
reveals that the tangential arc in A~370 has a redshift of $0.724$, in
good qualitative agreement with the requirement found here.

There is one potential difficulty with lensing by the profile (1.1)
though, because tangential arcs usually appear at distances of several
10 kpc from the cluster center, while numerical simulations yield
scale radii on the order of several 100 kpc. Therefore, the tangential
arcs must be formed well within the scale radius, and there their
radial magnification is high, $\sim3\ldots5$ for typical values. This
would imply that the radial source size of the arc sources should be
of order $0\farcs2\ldots0\farcs3$ in order to reconcile the observed
tangential-arc widths with this radial magnification.

\acknow{~I thank Peter Schneider, Simon White, and Jordi
Miralda-Escud\'e for useful comments. This work was supported in part
by the Sonderforschungsbereich SFB 375-95 of the Deutsche
Forschungsgemeinschaft.}

\begref{References}
\ref Bartelmann, M., Steinmetz, M., \& Weiss, A. 1995, A\&A, 297, 1
\ref Blandford, R.D., \& Narayan, R. 1992, ARA\&A, 30, 311
\ref Cole, S., \& Lacey, C. 1995, preprint
\ref Fort, B., Le F\`evre, O., Hammer, F., \& Cailloux, M. 1992, ApJ,
399, L125
\ref Hammer, F. 1991, ApJ, 383, 66
\ref Kneib, J.-P., Mellier, Y., Fort, B., \& Mathez, G. 1993, A\&A,
273, 367
\ref Lynds, R., \& Petrosian, V. 1989, ApJ, 336, 1
\ref Mellier, Y., Fort, B., \& Kneib, J.-P. 1993, ApJ, 407, 33
\ref Miralda-Escud\'e, J. 1993, ApJ, 403, 497
\ref Miralda-Escud\'e, J. 1995, ApJ, 438, 514
\ref Navarro, J., Frenk, C.S., \& White, S.D.M. 1995a, MNRAS, 275, 720
\ref Navarro, J., Frenk, C.S., \& White, S.D.M. 1995b, MNRAS, in press
\ref Schneider, P., Ehlers, J., \& Falco, E.E. 1992, Gravitational
Lenses (Heidelberg: Springer)
\ref Smail, I., Dressler, A., Kneib, J.-P., Ellis, R.S., Couch, W.J.,
Sharples, R.M., \& Oemler, A. 1995, SISSA preprint 9503063
\ref Soucail, G., Fort, B., Mellier, Y., \& Picat, J.P. 1987, A\&A,
172, L14
\ref Soucail, G., Mellier, Y., Fort, B., Mathez, G., \& Cailloux,
M. 1988, A\&A, 191, L19
\endref

\end